# Interface-Enhanced Superconductivity in Ultrathin TiN Proximitized by Topological Insulators

Renjie Xie,[1, #] Bowen Hao,[2, #] Min Ge,[3, #] Shenjin Zhang,[4, #] Rongjing Zhai,[1] Jiachang Bi,[1] Shunda Zhang,[1] Shaozhu Xiao,[5] Fengfeng Zhang,[4] Hee Taek Yi,[6] Seongshik Oh,[6] Tong Zhou,[2, *] Yanwei Cao,[1, *] and Xiong Yao[1, *]

[1]*Ningbo Institute of Materials Technology and Engineering, Chinese Academy of Sciences, Ningbo 315201, China*

[2]*Eastern Institute for Advanced Study, Eastern Institute of Technology, Ningbo 315200, China*

[3]*The Instruments Center for Physical Science, University of Science and Technology of China, Hefei 230026, China*

[4]*Key Laboratory of Functional Crystals and Laser Technology, Technical Institute of Physics and Chemistry (TIPC), Chinese Academy of Sciences, Beijing 100190, China.*

[5]*Yongjiang Laboratory, Ningbo 315202, China*

[6]*Department of Physics & Astronomy, Rutgers, The State University of New Jersey, Piscataway, New Jersey 08854, United States*

**# These authors contribute equally to this work**

**\*Email: tzhou@eitech.edu.cn**

**\*Email: ywcao@nimte.ac.cn**

**\*Email: yaoxiong@nimte.ac.cn**

# ABSTRACT

High-quality topological insulator-superconductor (TI-SC) heterostructure with an atomically sharp and well-controlled interface is crucial for realizing topological superconductivity and topological quantum qubit. In particular, many studies of TI-SC heterostructures have focused on inducing superconducting gap in the TI layer via proximity effect, while the active manipulation of superconductivity in the SC layer remains largely unexplored. In this work, we fabricated TI/TiN heterostructures using highly air-stable, ultrathin TiN films as the SC layer, and observed an interface-enhanced superconductivity that contrasts with the conventional proximity effect in superconductor-normal metal interface. Band structure measurements reveal a consistent shift of Dirac point with $T_c$ enhancement. Interfacial charge transfer provides a plausible explanation for this shift based on the systematic analysis and is therefore a likely contributor to the observed $T_c$ enhancement. First principles calculations elucidate the charge transfer pathways, highlighting the critical role of the interfacial BiTe (BiSe) bilayer. Our results not only provide a tunable TI-SC hybrid system with robust superconductivity at ultrathin thickness, but also offer a potential route for manipulating superconductivity in TI-SC heterostructures via interface engineering.

Superconductivity is a macroscopic quantum phenomenon characterized by zero electrical resistance below a critical temperature ($T_c$). Integrating superconductivity with topological materials can give rise to a topological superconductor, an exotic form of quantum matter predicted to host Majorana zero modes (MZMs). These quasiparticles, which obey non-Abelian statistics, are considered building blocks for realizing fault-tolerant quantum computing. An effective approach to implement topological superconductor involves coupling an s-wave superconductor with a topological insulator, forming a topological insulator-superconductor (TI-SC) heterostructure[1–3]. Various theoretical proposals for realizing MZMs, including those based on vortex cores at TI-SC interface or edge states in quantum anomalous Hall insulator-superconductor heterostructure[4–6], require the growth of high-quality TI-SC heterostructure with an atomically sharp and well-controlled interface.

Among various TI-SC heterostructures[7–17], the ones utilizing 3D TIs like $Bi_2Se_3$ or $Bi_2Te_3$ outperform other materials due to their spin non-degenerate helical surface states and a large band gap ~ 0.3 eV[13]. Compared to the readily obtained high-quality TI layer facilitated by van der Waals epitaxy, achieving a high-quality SC layer remains the major challenge in growing TI-SC heterostructures. First, most suitable superconducting materials are chemically reactive[7,10,18], and some of them tend to react with the TI layer, resulting in interfacial degradation or the formation of a dead layer[7]. This high reactivity also renders the SC layer air-sensitive and prone to surface degradation, requiring stringent *in-situ* growth[10,12,13,19–21] or *in-situ* cleaving processes[8,9,22–24]. Furthermore, the superconductivity in most superconducting thin films deteriorates rapidly as the thickness is reduced to atomic scale, often leading to a broadened superconducting transition or even complete loss of superconductivity[13,19,25,26]. Due to the lack of ultrathin SC films with sharp transition, prevailing studies focus on inducing superconducting gaps in TI layer[7,8,27,28], while the reverse effect, i.e., how to manipulate the superconductivity in SC layer via proximity coupling with TI layer, remains largely unexplored. The shunting effect from the thick SC layer renders the subtle interactions at the TI-SC interface very challenging to detect by transport measurements, obscuring the understanding of interactions between SC and TI layers at the atomically thin interface.

In this work, we introduce *ex situ* grown ultrathin TiN (111) film as the desired SC layer that demonstrates extraordinary environmental stability[29–32]. The 6-fold structural symmetry of TiN (111) enables high compatibility with TIs, resulting in superior interfacial quality. Compared to superconductors like $Bi_2Sr_2CaCu_2O_{8+\delta}$[33], $Nd_{0.8}Sr_{0.2}NiO_2$[34], $NbSe_2$[35], and $FeSe$[36], which are prone to degradation in air, however, the robust TiN surface is resistant to degradation, contamination, and even acid or alkali corrosion[37–40]. We used a magnetron sputtering epitaxy system to grow ultrathin TiN films as SC layers and transfer them to molecular beam epitaxy (MBE) chamber to fabricate TI/TiN heterostructures. We demonstrate that the TiN films preserve robust superconductivity with sharp transitions at ultrathin thickness ~ 4 nm, thereby minimizing bulk effects and enabling clear probing of variation in superconductivity. Most surprisingly, we observed an enhancement of superconducting temperature $T_c$ in $Bi_2Te_3(Bi_2Se_3)$/TiN heterostructures compared to pure TiN film. This phenomenon contrasts sharply with the conventional superconducting proximity effect at a superconductor-normal metal interface, where the proximity coupling quenches the superconductivity and lead to a $T_c$ reduction in the SC layer[41,42]. The origin of the enhanced superconductivity is investigated by transport measurements, angle-resolved photoemission spectroscopy (ARPES), and first-principles calculations, providing valuable insights on interfacial manipulation of superconductivity in TI-SC system.

## RESULTS

**Structural characterizations of the TI/TiN heterostructures**

Figure 1a depicts the schematic illustration of the TI ($Bi_2Te_3$ or $Bi_2Se_3$)/TiN heterostructures. Notably, a cubic BiTe (BiSe) bilayer naturally formed at the interface between TI and TiN layers during the growth, which is characterized by the scanning transmission electron microscopy (STEM) imaging in Figure 1d and e. $Bi_2Te_3$ and $Bi_2Se_3$ are benchmark 3D TIs with a single Dirac cone located at the Γ point of the first Brillouin zone[43]. Their unit cells are composed of quintuple-layer (QL) blocks with a Te(Se)–Bi–Te(Se)–Bi–Te(Se) atomic arrangement. Figure 1b gives the X-ray diffraction (XRD) 2θ scans for both the $Bi_2Te_3$/TiN and $Bi_2Se_3$/TiN heterostructures, which clearly show the $Bi_2Te_3$ (00 3n), $Bi_2Se_3$ (00 3n), and TiN (mmm) peaks, without any sign of secondary phases. The rocking curve scans in Figure S1 yield full width at half maximum (FWHM) values of approximately 0.1° for the $Bi_2Te_3$/TiN heterostructure and 0.06° for the $Bi_2Se_3$/TiN heterostructure, respectively. The XRD results together with the distinct low energy electron diffraction (LEED) patterns (Figure S4) indicate the high-quality epitaxy for TI/TiN heterostructures, presumably owing to the same sixfold symmetry between TiN(111) and $Bi_2Te_3$ ($Bi_2Se_3$)[37].

Heterostructures grown on *ex-situ* films typically exhibit degraded surface quality or even interfacial dead layers. However, the STEM images (Figure 1c-e and Figure S5) of both $Bi_2Te_3$/TiN and $Bi_2Se_3$/TiN heterostructures grown on *ex-situ* TiN films (as thin as 4.2±0.2 nm) demonstrate atomically sharp and well-defined interfaces. Figure 1d shows an enlarged view of the $Bi_2Te_3$/TiN interface, revealing a distinct bilayer structure (~3.3 Å thick) that differs from the $Bi_2Te_3$ phase. Based on a careful analysis of the atomic spacing (Figure S6), we identify this bilayer structure as the cubic BiTe phase. A similar cubic BiSe bilayer is also observed at the $Bi_2Se_3$/TiN interface (Figure 1e) and has previously been reported at the $Bi_2Se_3/NbSe_2$ interface[12]. The atomic force microscopy (AFM) morphology of TiN film exhibits a surface roughness comparable to that of commercial $Al_2O_3$ substrates[37] (146 pm), providing an atomically flat surface for the epitaxial growth of TI layers. The AFM images of both $Bi_2Te_3$/TiN (Figure S8) and $Bi_2Se_3$/TiN (Figure 1f) show characteristic triangular-shaped terraces with height ~ 1 nm, corresponding to 1 QL of $Bi_2Te_3$ or $Bi_2Se_3$. These large, flat terraces suggest a good structural match between the TI layers and TiN surface[37], which is in agreement with the XRD and STEM results.

**Enhanced superconductivity in TI/TiN heterostructures**

We performed *ex situ* electrical transport measurements on both pure TiN films and TI/TiN heterostructures to evaluate their superconducting properties. We note that the superconducting transition temperature $T_c$ of TiN films varies across different sputtering batches. Therefore, all data presented within a single figure are grown on TiN films from the same sputtering batch to ensure a rigorous comparison. Figure 2a compares the temperature-dependent normalized longitudinal resistance of a pure TiN film and $Bi_2Te_3$/TiN heterostructures with varying thicknesses of $Bi_2Te_3$ (8, 12, and 24 nm), all fabricated on 4 nm TiN films from the same batch. The 4 nm ultrathin TiN film exhibits a sharp superconducting transition with $T_c$ (defined as the temperature at which R drops to 50% of the normal-state resistance measured at 7.0 K) of 4.3 K, demonstrating no degradation of superconductivity even at near-atomic scale. After the deposition of $Bi_2Te_3$ on TiN surface, a noticeable enhancement in $T_c$ is observed for the n=8 heterostructure, and the enhancement become more pronounced in the n = 12 and 24 heterostructures. The n=12 sample reaches a maximal $T_c$ of 4.9 K, about 0.6 K higher than that of pure TiN. In contrast, Figure 2c shows a 12nm $Bi_2Te_3$/4nm TiN control sample with a deliberately deteriorated interface lacking the BiTe bilayer, which exhibits a $T_c$ reduction of 0.6 K (Figure 2d). This result unambiguously validates that high quality interface with BiTe bilayer is critical for the superconductivity enhancement.

Then we replaced $Bi_2Te_3$ with $Bi_2Se_3$ and grow $Bi_2Se_3$/TiN (Figure 2b) heterostructures, in which the $T_c$ of pure TiN was measured as 4.2 K. The $T_c$ of the $Bi_2Se_3$/TiN heterostructures increased gradually from n = 4 to n = 24, reaching a maximal value in the n=24 sample with an enhancement of ~ 0.5 K compared to pure TiN. As a comparison, Figure 2e shows the resistance of a $Bi_2Se_3$/TiN heterostructure grown on thick TiN (70 nm), exhibiting a nearly identical $T_c$ (change <0.1 K) with pure TiN. Notably, STEM measurements reveal that the cubic BiSe interfacial phase is absent in this sample (Figure S7), in contrast to the $Bi_2Se_3$/4 nm TiN case. Notably, STEM measurements reveal that the cubic BiSe interfacial phase is absent in this sample (Figure S7), in contrast to the $Bi_2Se_3$/4 nm TiN case. This structural difference is likely due to the rougher TiN surface with higher nucleation sites in thick TiN (Figure S8), which favors direct van der Waals $Bi_2Se_3$ nucleation over the metastable BiSe phase. Consequently, this structural difference provides a plausible explanation for the absence of $T_c$ enhancement and likely reflects the distinct interfacial conditions in thick TiN films. This observation implies that a smoother TiN surface with lower nucleation density may be a necessary condition for BiSe interfacial layer formation. Next, we replaced the TI layer with a non-topological material, Se, as shown in Figure 2f. The resistance curve of the 140nm Se/4nm TiN sample closely resembles that of the pure TiN film, while the 50nm $Bi_2Se_3$/4nm TiN heterostructure shows a clear $T_c$ enhancement of 0.4 K, indicating that the TI layer could plays an essential role in enhancing superconductivity.

**Thickness-dependent evolution of band structure**

Next, we performed *in situ* ARPES measurements on $Bi_2Te_3$/TiN heterostructures with varying $Bi_2Te_3$ thickness. The band structure of ultrathin pure TiN film is difficult to resolve due to the requirement of *ex situ* transfer, considering ARPES is a surface sensitive probe. Figure S9 gives the hexagonal Fermi surface mapping of a 50nm $Bi_2Te_3$/4nm TiN heterostructure, which is consistent with previous study on $Bi_2Te_3$[44]. Figure 3 presents the ARPES spectra and corresponding energy distribution curves collected along the $\bar{\Gamma}-\bar{K}$ direction for $Bi_2Te_3$/4nm TiN heterostructures with $Bi_2Te_3$ thickness ranging from 1 nm to 24 nm. For the 1 nm $Bi_2Te_3$/TiN sample (n=1), the ARPES spectra exhibits a nearly parabolic band dispersion, similar to that of 1QL $Bi_2Te_3$ grown on Si(111) substrate[45]. In the n=2 sample, a quantum-well (QW) state appears in the bulk conduction band, indicating that the film thickness is within the quantum confinement regime. As the thickness increases, surface states form a gapless Dirac cone, causing the quantum well states to gradually disappear, overshadowed by the spectral weight of the topological surface states. Additionally, a new valence band emerges at $E-E_F \approx -0.45$ eV in the n=2 sample, resulting in a gap of approximately 0.13 eV to the conduction band. As the thickness of the $Bi_2Te_3$ layer increases, these two bands evolve into a gapless surface state, forming a typical Dirac cone. Meanwhile, the conduction and valence bands cross at $E-E_F \approx -0.32$ eV, with the crossover thickness determined to be 3 nm, consistent with previous studies[45]. For samples with n≥3, the Dirac point gradually shifts toward the Fermi level with increasing thickness until it reaches a saturation value at approximately 24 nm. Although ARPES is surface sensitive and does not directly probe the $Bi_2Te_3$/TiN interface, it detects interface-induced modifications of the surface electronic structure. Interfacial charge transfer alters the surface chemical potential via band bending and charge redistribution, manifesting as a shift of the Dirac point at the surface. Figure 5a summarizes the thickness-dependent shift of Dirac point in $Bi_2Te_3$/TiN alongside the corresponding $T_c$ enhancement in both Bi$_2$Te$_3$/TiN and Bi$_2$Se$_3$/TiN heterostructures. While the Dirac point in TI films can evolve with thickness due to intrinsic vacancies and antisite defects, control experiments on Bi$_2$Te$_3$ films grown on InP (Figure S10-11) under identical conditions show a much smaller maximum shift (~60 meV) than the ~100 meV shift observed in Bi$_2$Te$_3$/TiN heterostructures. Consistent with this conclusion, previous ARPES studies on Bi$_2$Te$_3$ films grown on Si substrates also report only a relatively weak thickness-dependent Dirac point shift that is much smaller than 100 meV[45]. Moreover, in Bi$_2$Te$_3$/TiN the pronounced downward shift of the Dirac point from 12 to 24 QL closely resembles the change in $T_c$ observed in transport measurements, whereas such a downward shift is absent in the Bi$_2$Te$_3$/InP control samples. Together, these contrasts indicate an additional interface induced contribution in the Bi$_2$Te$_3$/TiN system and suggest that intrinsic bulk defects alone cannot account for the ARPES trends, making interfacial charge transfer a more plausible explanation.

To better understand the mechanism of our observations, we carried out first-principles calculations for the Bi$_2$Te$_3$/BiTe/TiN heterostructure on a seven-layer TiN(111) slab (Figure 4a-b). The charge-density difference indicates that charge transfer is dominated by the BiTe-TiN interface, with a smaller contribution at the Bi$_2$Te$_3$-BiTe interface. At the Bi$_2$Te$_3$-BiTe interface (Figure 4c), pronounced electron-density overlap between Te (Bi$_2$Te$_3$) and Bi/Te (BiTe) reveals covalent-like electron sharing that establishes an interfacial charge-transfer pathway. Electron accumulation within Bi sites away from this interface further suggests donation from the Bi$_2$Te$_3$ side and/or intra-layer rearrangement in BiTe. These features are consistent with the planar-averaged charge-density difference along the z direction (right panel of Figure 4c). In contrast, the BiTe-TiN interface (Figure 4d) exhibits strong charge rearrangement characterized by electron depletion near Bi and accumulation at N, evidencing direct electron transfer from Bi to the more electronegative N. The corresponding planar-averaged profile shows a negative peak at Bi and a characteristic double positive peak at N, arising from the two-dimensional averaging across distinct charge regions. Electron-localization-function analysis (Figure 4e) points to predominantly ionic Bi–N bonding, consistent with the charge-density-difference results. The states near $E_F$ are dominated by Ti 3d character, and the interfacial Ti layers exhibit an enhanced $N(E_F)$ relative to pristine TiN (Figure 4f), consistent with interfacial hybridization and charge transfer.

We also considered both surface terminations of TiN (Ti- versus N-terminated) as limiting boundary conditions for the $Bi_2Se_3$-BiSe-TiN in our calculations (Figure S12). Work function analysis reveals a thermodynamic driving force for electron transfer from the TI side into TiN for both terminations, with a larger magnitude for N termination. Differential charge-density maps show robust, interface-localized charge redistribution in both cases, indicating that interfacial charge transfer is enabled by the work-function mismatch and is robust against the choice of surface termination, with the termination primarily modulating its strength. Layer-resolved band projections for $Bi_2Se_3$/TiN heterostructure identify the TI and TiN contributions near $E_F$ and visualize the interface coupling (Figure S13). To ensure clear interpretability, we employed a minimal commensurate interface model without BiSe layer that avoids severe band folding while preserving the key features of charge transfer into TiN and the associated enhancement of Ti-derived states near $E_F$. We note that the $Bi_2Se_3$ projected feature near $\Gamma$ point appears low dispersion ("flat-like"), which likely originates from interface induced hybridization and may enhance local DOS, potentially relevant to the superconductivity enhancement.

**Studies on upper critical fields**

In order to investigate the dimensionality of superconductivity in $Bi_2Te_3$/TiN heterostructures, we measured the magnetic field dependent resistance with magnetic fields applied along perpendicular and parallel to the ab plane (Figure S15-16). The upper critical fields ($H_{c2,\perp}$ and $H_{c2,\parallel}$), defined by the point where the resistance drops to 50% of the normal-state resistance, are shown in Figure 5b-c. When an external magnetic field is applied to a superconductor, the upper critical field $H_{c2}$ is determined by two main effects[12,46,47]. The orbital effect arises from the pair breaking induced by the coupling between electron momentum and magnetic field, exhibiting a strong temperature dependence that can be described by Ginzburg-Landau (GL) theory:

$$H_{C2}^{\perp}(T) = \frac{\Phi_0}{2\pi\xi_{GL}(0)^2}\left(1-\frac{T}{T_C}\right), \tag{1}$$

$$H_{C2}^{\parallel}(T) = \frac{\sqrt{3}\Phi_0}{\pi\xi_{GL}(0)d}\left(1-\frac{T}{T_C}\right)^{1/2}, \tag{2}$$

where $\Phi_0$ is the flux quantum, d is the superconducting thickness, and $\xi_{GL(0)}$ is the zero-temperature GL coherence length. The Pauli paramagnetic effect results from the electron spin alignment by magnetic field and establishes a Pauli limit of $H_p$ ($H_p \approx 1.86\ T_c$). The orbital-limiting effect typically dominates pair breaking at temperatures near $T_c$, whereas the Pauli paramagnetic mechanism becomes predominant at low temperatures far away from $T_c$. Figure 5b gives the reduced-temperature ($T/T_c$) dependence of $H_{c2,\perp}$, exhibiting a clear linear behavior in both pure TiN film and $Bi_2Te_3$/TiN heterostructures. For the pure TiN (n=0) sample, the $H_{c2,\perp}$(T=0 K) is 1.3 T. As $Bi_2Te_3$ increases, $H_{c2,\perp}$ first increases to a maximum of 1.6 T at n=12 and then slightly decreases at n=24, following the same trend observed in $T_c$ enhancement. The inset of Figure 5b exhibits the extracted GL coherence length $\xi_{GL,\perp}$ extracted by fitting the $H_{c2,\perp}$ using Eq. 1, which decreases from 15.7 nm in pure TiN to a minimum of 14.4 nm at n=12, and then slightly increases at n=24. Notably, in superconductor–normal metal heterostructures, the superconducting proximity effect typically leads to an increase in the coherence length[12]; however, in ultrathin TiN proximitized by $Bi_2Te_3$, the coherence length decreases instead, indicative of an unconventional underlying mechanism.

Figure 5c presents the $H_{c2,//}$ for pure TiN and the 12nm $Bi_2Te_3$/4nm TiN heterostructure with the highest $T_c$. According to GL theory, 2D superconductors exhibit a square root temperature dependence $H_{c2,//} \propto \sqrt{(1 - T/T_c)}$, while strongly coupled 3D-like superconductors follow a linear behavior of $[H_{c2,//} \propto (1 - T/T_c)]$[48]. As shown in Figure 5c, the $H_{c2,//}$ for both pure TiN and $Bi_2Te_3$/TiN is well described by a linear 3D dependence near $T_c$, then crossing over to a 2D square root dependence at lower temperatures. This 3D to 2D crossover behavior highly resembles the observations in quasi-two-dimensional Nb/Ge composites[48]. The dimensional crossover is also corroborated by the fitting models with an unfixed exponent in Figure S17b, which gives a fitting exponent of 0.78 lies between 1 and 0.5. Using the 2D model at low temperatures, we extrapolated $H_{c2,//}(T = 0 K)$ to be 7.8 T for pure TiN and 8.8 T for 12nm $Bi_2Te_3$/TiN. In 2D superconductors under in-plane magnetic fields, the orbital pair breaking becomes less effective and the Pauli paramagnetic effect dominates as the temperature decreases. We calculated the Pauli limit of the upper critical field to be 8.0 T for pure TiN and 9.1 T for 12nm $Bi_2Te_3$/TiN heterostructure, highly consistent with the values derived from 2D GL theory, confirming the Pauli dominated pair breaking and 2D superconductivity at low temperatures.

**DISCUSSION**

Based on the data presented in Figure 2a and 2b, we observed a clear thickness-dependent $T_c$ enhancement in TI/TiN heterostructures. One might attribute the enhanced superconductivity to strain effects, which is more pronounced in ultrathin TiN than in thick TiN. However, in our heterostructures, the TI layer is grown via van der Waals epitaxy that is believed not to introduce substantial lattice strain into the underlying TiN layer. HRXRD measurements (Figure S2a-b) show that the TiN(111) peak position remains unchanged within experimental resolution after the growth of $Bi_2Se_3$ on 70 nm TiN and $Bi_2Te_3$ on 4 nm TiN, indicating negligible out-of-plane lattice distortion. Consistently, reciprocal space mapping (Figure S2c) yields layer spacings of $d_{(111)} \approx 2.450$Å and $d_{(11\bar{2})} \approx 1.728$Å, in excellent agreement with reported values for bare TiN[39]. Furthermore, STEM-HAADF analysis (Figure S2d and Figure S3d-f) reveals that the in-plane lattice parameter $a$ in both $Bi_2Te_3$/4 nm TiN and $Bi_2Se_3$/4 nm TiN heterostructures matches bare 4nm TiN with noise level deviations < 1%. Together, these results indicate that strain effects are negligible in both heterostructures. Moreover, the $T_c$ enhancement is absent in both Se/4nm TiN and $Bi_2Te_3$/4nm TiN with deteriorated interface, unambiguously confirming that the superconductivity enhancement is not related to strain effect. To summarize the observations in Figure 2, the superconductivity enhancement is exclusively observed at atomically sharp and well-defined interfaces formed by an ultrathin TiN layer and a TI layer incorporating a BiTe (BiSe) bilayer. These results are in stark contrast to the proximity coupling in a superconductor-normal metal interface, indicative of an underlying mechanism distinct from the conventional superconducting proximity effect[41,42]. Figure 5a shows a clear correlation between the thickness-dependent shift of the Dirac point and the $T_c$ enhancement, consistent with interfacial charge transfer as a likely contributor to the superconductivity enhancement. Comparative analysis of the differential-charge maps at $Bi_2Te_3$–BiTe and BiTe–TiN interfaces in Figure 4 reveals two complementary transfer mechanisms: electronegativity-driven, directional injection at BiTe (BiSe)–TiN, and covalent-like electron sharing at $Bi_2Te_3(Bi_2Se_3)$–BiTe(BiSe). Their synergy forms a continuous transfer channel, with the BiTe(BiSe) layer serving as the critical bridge that mediates efficient electron flow from $Bi_2Te_3(Bi_2Se_3)$ to TiN.

A recent ARPES study indicate that TiN exhibits a sizable Coulomb interaction (U∼6.2–8.5 eV)[39], placing it closer to a strongly correlated regime rather than a weak-coupling superconductor. In this context, we do not interpret the interfacial charge transfer in a carrier doping picture. Instead, the charge redistribution at the $Bi_2Te_3$/TiN interface probably induces a local electronic reconstruction in the interfacial TiN layers by modifying electronic screening, the effective Coulomb interaction $U_{eff}$, and the local bandwidth W. Such changes renormalize the ratio $U_{eff}/W$, which can shift the system toward a more favorable correlation regime for pairing.

**CONCLUSIONS**

In summary, we fabricated $Bi_2Te_3$/TiN and $Bi_2Se_3$/TiN heterostructures and observed interface-enhanced superconductivity, which occurs exclusively at atomically sharp TI/TiN interfaces featuring an intercalated BiTe (BiSe) bilayer. The ARPES measurements are consistent with interfacial charge transfer as a likely contributor to the observed superconductivity enhancement. First-principles calculations further elucidate the detailed electron transfer pathways, in which the interfacial BiTe bilayer plays a critical role. This work shows that TI/TiN heterostructure is a promising and tunable TI-SC hybrid system, as the TiN layer eliminates the need for stringent *in situ* growth and maintains sharp superconducting transitions even at ultrathin thicknesses. More importantly, our results offer an effective strategy to manipulate superconductivity in TI-SC heterostructures via interface engineering, i.e. enhancing it with BiTe interfacial layer or suppressing it by deliberately controlled interface quality, facilitating the search for novel topological superconducting states. At last, our work unveils a new TI-SC family composed of nitride superconductors and topological materials, as a promising candidate for interface-enhanced superconductivity, providing an opportunity for exploring this effect in higher-$T_c$ materials such as NbN or NbTiN ($T_c > 15$ K).

**METHODS**

**Film growth.** Epitaxial TiN (111) films were grown on 5×5 mm² α-Al$_2$O$_3$ (0001) substrates by a homemade magnetron sputtering epitaxy system, the details of which were reported in previous reports[38,39]. The films in Figure 2b were grown on 5×5 mm² KTaO$_3$ (111) substrates. Subsequently, the as-grown TiN films were taken out and transferred into an MBE chamber and annealed at 700 °C under ultra-high vacuum (UHV) for 12 hours to remove surface contaminants. High-purity Bi (99.997%), Se (99.999%), and Te (99.9999%) were evaporated from Knudsen effusion cells. The TI growth followed the protocols described in our previous report[37]. The growth rate was set as ~1 QL/min for the TI films, which was calibrated in situ by quartz crystal microbalance and X-ray reflectivity. Notably, we observed a clear $T_c$ enhancement in Bi$_2$Se$_3$/TiN heterostructures grown on both KTaO$_3$ (Figure 2b) and Al$_2$O$_3$ (Figure 2f) substrates.

**Crystal Structure and Surface Analysis.** Low-energy electron diffraction images were collected *in situ* after the sample growth. The crystal structure was characterized by high-resolution X-ray diffraction (HRXRD, Bruker D8 Discovery) measurements with Cu Kα radiation ($\lambda = 1.5406$ Å) as the X-ray source. Symmetrical θ–2θ measurements, rocking curves, and X-ray reflectivity were performed with 40 kV working voltage and 40 mA current. The surface morphology was examined by a Bruker Dimension Icon AFM with Nanoscope V controller (Digital Instruments, USA). All topographic images were acquired in scanAsyst mode with silicon cantilevers. The scanning transmission electron microscopy (STEM) samples were prepared by using focused ion beam (FIB) milling. High-angle annular dark-field scanning transmission (HAADF-STEM) and energy-dispersive X-ray (EDX) spectroscopy measurements were conducted on a Themis Z Double spherical aberration corrected transmission electron microscope with 300 kV.

**Transport Measurements.** All the transport measurements were carried out in a Quantum Design physical properties measurement system (PPMS) with a base temperature of 1.8 K and a magnetic field up to 9 T. Electrical electrodes were made by manually pressing four indium wires on each sample.

**Angle-Resolved Photoemission Spectroscopy Measurements**. The electronic structures of $Bi_2Te_3$/TiN heterostructures were measured at 7.3 K in a homemade angle-resolved photoemission spectroscopy (ARPES) system with a base pressure of ~$5\times10^{-11}$ mbar. The photoelectrons were excited using a 177 nm deep-ultraviolet laser ($h_v$ = 6.997 eV) with a total energy resolution better than 1 meV. The Fermi level was determined by the measurement of a polycrystalline gold sample in electrical contact with itself.

**First-principles calculations.** All the calculations were carried out using the Vienna Ab initio Simulation Package (VASP)[49,50] with the Projector Augmented Wave (PAW) method. Exchange and correlation effects were treated within the generalized gradient approximation (GGA) using the Perdew–Burke–Ernzerhof (PBE) functional[51]. The Brillouin zone was sampled with a $5 \times 5 \times 1$ Gamma-centered Monkhorst–Pack k-point grid. A plane-wave cutoff energy of 500 eV was applied. All atomic positions were fully relaxed until the energy and forces converged to $10^{-5}$ eV and 0.01 eV/Å, respectively. A vacuum layer of 20 Å was added along the z-direction to avoid interactions between periodic images. To achieve lattice matching with BiTe, we expanded TiN into a $3\sqrt{3} \times 5 \times 1$ supercell and used its lattice parameters as the reference. Accordingly, BiTe was expanded into a $\sqrt{5} \times \sqrt{5} \times 1$ supercell, and $Bi_2Te_3$ into a $2\sqrt{3} \times 3 \times 1$ supercell. This configuration introduced no strain to the TiN layer, while the strain in both BiTe and $Bi_2Te_3$ was kept below 5%, ensuring the accuracy of the calculations. The differential charge was defined as $\Delta\rho = \rho(all) - \rho(TiN) - \rho(BiTe) - \rho(Bi_2Te_3)$. We also used $\Delta\rho = \rho(all) - \rho(TiN) - \rho(BiTe/Bi_2Te_3)$ as an alternative approach for verification. The results showed no substantial difference (Figure S14), confirming the robustness of our calculations.

## SUPPLEMENTARY INFORMATION

The Supporting Information is available free of charge.

The HRXRD characterizations, LEED pattern, STEM images and analysis, AFM images, ARPES data, DFT calculations and the transport property data.

## ACKNOWLEDGMENTS

The work is supported by the National Natural Science Foundation of China (Grant Nos. 12304541, 12474155), the National Key Research and Development Program of China (Grant Nos. 2024YFF0508500 and 2022YFA1403000), the Hundred Talents Program of Chinese Academy of Sciences, the Zhejiang Provincial Natural Science Foundation of China (LR25A040001, LRG25E020001), and the Ningbo Science and Technology Bureau (Grant Nos. 2023J047). The computational resources for this research were provided by the High Performance Computing Platform at the Eastern Institute of Technology, Ningbo.

## AUTHOR CONTRIBUTIONS

R.X., B.H., M.G. and Shenjin Z. contribute equally to this work. X.Y. and Y.C. conceived the experiments. R.X., X.Y. grew the heterostructures with the help of H.Y. and S.O. R. Z, J.B., Shunda Z. and Y.C. grew the TiN films. M.G. performed the STEM measurements and analyzed the data with R.X. R.X., Shenjin Z., F.Z. and S.X. performed the ARPES measurements. R.X. and X.Y. performed the transport measurements. B.H. and T.Z. performed all the calculations. R.X., B.H, T.Z. and X.Y. analyzed the results and wrote the manuscript with contributions from all authors.

## COMPETING INTERESTS

The authors declare no competing financial interest.

## DATA AVAILABILITY

The data that support the findings of this study are available from the corresponding authors upon reasonable request.

## REFERENCES

(1) Lutchyn, R. M.; Sau, J. D.; Das Sarma, S. Majorana Fermions and a Topological Phase Transition in Semiconductor-Superconductor Heterostructures. *Phys. Rev. Lett.* **2010**, *105* (7), 077001. https://doi.org/10.1103/PhysRevLett.105.077001.
(2) Oreg, Y.; Refael, G.; Von Oppen, F. Helical Liquids and Majorana Bound States in Quantum Wires. *Phys. Rev. Lett.* **2010**, *105* (17), 177002. https://doi.org/10.1103/PhysRevLett.105.177002.
(3) Fu, L.; Kane, C. L. Superconducting Proximity Effect and Majorana Fermions at the Surface of a Topological Insulator. *Phys. Rev. Lett.* **2008**, *100* (9), 096407. https://doi.org/10.1103/PhysRevLett.100.096407.
(4) Sato, M.; Ando, Y. Topological Superconductors: A Review. *Rep. Prog. Phys.* **2017**, *80* (7), 076501. https://doi.org/10.1088/1361-6633/aa6ac7.
(5) Qi, X.-L.; Hughes, T. L.; Zhang, S.-C. Chiral Topological Superconductor from the Quantum Hall State. *Phys. Rev. B* **2010**, *82* (18), 184516. https://doi.org/10.1103/PhysRevB.82.184516.
(6) Chung, S. B.; Qi, X.-L.; Maciejko, J.; Zhang, S.-C. Conductance and Noise Signatures of Majorana Backscattering. *Phys. Rev. B* **2011**, *83* (10), 100512. https://doi.org/10.1103/PhysRevB.83.100512.

(7) Flötotto, D.; Ota, Y.; Bai, Y.; Zhang, C.; Okazaki, K.; Tsuzuki, A.; Hashimoto, T.; Eckstein, J. N.; Shin, S.; Chiang, T.-C. Superconducting Pairing of Topological Surface States in Bismuth Selenide Films on Niobium. *Sci. Adv.* **2018**, *4* (4), eaar7214. https://doi.org/10.1126/sciadv.aar7214.

(8) Wang, E.; Ding, H.; Fedorov, A. V.; Yao, W.; Li, Z.; Lv, Y.-F.; Zhao, K.; Zhang, L.-G.; Xu, Z.; Schneeloch, J.; Zhong, R.; Ji, S.-H.; Wang, L.; He, K.; Ma, X.; Gu, G.; Yao, H.; Xue, Q.-K.; Chen, X.; Zhou, S. Fully Gapped Topological Surface States in $Bi_2Se_3$ Films Induced by a D-Wave High-Temperature Superconductor. *Nat Phys* **2013**, *9* (10), 621–625. https://doi.org/10.1038/nphys2744.

(9) Wang, M.-X.; Liu, C.; Xu, J.-P.; Yang, F.; Miao, L.; Yao, M.-Y.; Gao, C. L.; Shen, C.; Ma, X.; Chen, X.; Xu, Z.-A.; Liu, Y.; Zhang, S.-C.; Qian, D.; Jia, J.-F.; Xue, Q.-K. The Coexistence of Superconductivity and Topological Order in the $Bi_2Se_3$ Thin Films. *Science* **2012**, *336* (6077), 52–55. https://doi.org/10.1126/science.1216466.

(10) Yang, H.; Li, Y.; Liu, T.; Xue, H.; Guan, D.; Wang, S.; Zheng, H.; Liu, C.; Fu, L.; Jia, J. Superconductivity of Topological Surface States and Strong Proximity Effect in $Sn_{1-x}Pb_xTe$–Pb Heterostructures. *Adv. Mater.* **2019**, *31* (52), 1905582. https://doi.org/10.1002/adma.201905582.

(11) Hlevyack, J. A.; Najafzadeh, S.; Lin, M.-K.; Hashimoto, T.; Nagashima, T.; Tsuzuki, A.; Fukushima, A.; Bareille, C.; Bai, Y.; Chen, P.; Liu, R.-Y.; Li, Y.; Flötotto, D.; Avila, J.; Eckstein, J. N.; Shin, S.; Okazaki, K.; Chiang, T.-C. Massive Suppression of Proximity Pairing in Topological $(Bi_{1-x}Sb_x)_2Te_3$ Films on Niobium. *Phys. Rev. Lett.* **2020**, *124* (23), 236402. https://doi.org/10.1103/PhysRevLett.124.236402.

(12) Yi, H.; Hu, L.-H.; Wang, Y.; Xiao, R.; Cai, J.; Hickey, D. R.; Dong, C.; Zhao, Y.-F.; Zhou, L.-J.; Zhang, R.; Richardella, A. R.; Alem, N.; Robinson, J. A.; Chan, M. H. W.; Xu, X.; Samarth, N.; Liu, C.-X.; Chang, C.-Z. Crossover from Ising- to Rashba-Type Superconductivity in Epitaxial $Bi_2Se_3$/Monolayer $NbSe_2$ Heterostructures. *Nat. Mater.* **2022**, *21* (12), 1366–1372. https://doi.org/10.1038/s41563-022-01386-z.

(13) Yi, H.; Hu, L.-H.; Zhao, Y.-F.; Zhou, L.-J.; Yan, Z.-J.; Zhang, R.; Yuan, W.; Wang, Z.; Wang, K.; Hickey, D. R.; Richardella, A. R.; Singleton, J.; Winter, L. E.; Wu, X.; Chan, M. H. W.; Samarth, N.; Liu, C.-X.; Chang, C.-Z. Dirac-Fermion-Assisted Interfacial Superconductivity in Epitaxial Topological-Insulator/Iron-Chalcogenide Heterostructures. *Nat Commun* **2023**, *14* (1), 7119. https://doi.org/10.1038/s41467-023-42902-2.

(14) He, Q. L.; Liu, H.; He, M.; Lai, Y. H.; He, H.; Wang, G.; Law, K. T.; Lortz, R.; Wang, J.; Sou, I. K. Two-Dimensional Superconductivity at the Interface of a $Bi_2Te_3$/FeTe Heterostructure. *Nat Commun* **2014**, *5* (1), 4247. https://doi.org/10.1038/ncomms5247.

(15) Chen, A.-H.; Lu, Q.; Hershkovitz, E.; Crespillo, M. L.; Mazza, A. R.; Smith, T.; Ward, T. Z.; Eres, G.; Gandhi, S.; Mahfuz, M. M.; Starchenko, V.; Hattar, K.; Lee, J. S.; Kim, H.; Moore, R. G.;

Brahlek, M. Interfacially Enhanced Superconductivity in Fe(Te,Se)/Bi$_4$Te$_3$ Heterostructures. *Adv. Mater.* **2024**. *36*, 2401809. https://doi.org/10.1002/adma.202401809.

(16) Moore, R. G.; Lu, Q.; Jeon, H.; Yao, X.; Smith, T.; Pai, Y.; Chilcote, M.; Miao, H.; Okamoto, S.; Li, A.; Oh, S.; Brahlek, M. Monolayer Superconductivity and Tunable Topological Electronic Structure at the Fe(Te,Se)/Bi$_2$Te$_3$ Interface. *Adv. Mater* **2023**, *35* (22), 2210940. https://doi.org/10.1002/adma.202210940.

(17) Brahlek, M.; Lapano, J.; Lee, J. S. Topological Materials by Molecular Beam Epitaxy. *J.Appl. Phys* **2020**, *128* (21), 210902. https://doi.org/10.1063/5.0022948.

(18) Sochnikov, I.; Bestwick, A. J.; Williams, J. R.; Lippman, T. M.; Fisher, I. R.; Goldhaber-Gordon, D.; Kirtley, J. R.; Moler, K. A. Direct Measurement of Current-Phase Relations in Superconductor/Topological Insulator/Superconductor Junctions. *Nano Lett.* **2013**, *13* (7), 3086–3092. https://doi.org/10.1021/nl400997k.

(19) Yi, H.; Zhao, Y.-F.; Chan, Y.-T.; Cai, J.; Mei, R.; Wu, X.; Yan, Z.-J.; Zhou, L.-J.; Zhang, R.; Wang, Z.; Paolini, S.; Xiao, R.; Wang, K.; Richardella, A. R.; Singleton, J.; Winter, L. E.; Prokscha, T.; Salman, Z.; Suter, A.; Balakrishnan, P. P.; Grutter, A. J.; Chan, M. H. W.; Samarth, N.; Xu, X.; Wu, W.; Liu, C.-X.; Chang, C.-Z. Interface-Induced Superconductivity in Magnetic Topological Insulators. *Science* **2024**, *383* (6683), 634–639. https://doi.org/10.1126/science.adk1270.

(20) Chang, K.; Hu, M.; Lin, H.; Liu, J.; Xue, Q.-K.; Chen, X.; Ji, S.-H. Oscillation of Electronic-Band-Gap Size Induced by Crystalline Symmetry Change in Ultrathin PbTe Films. *Phys. Rev. Lett.* **2023**, *131* (1), 016202. https://doi.org/10.1103/PhysRevLett.131.016202.

(21) Chang, K.; Deng, P.; Zhang, T.; Lin, H.-C.; Zhao, K.; Ji, S.-H.; Wang, L.-L.; He, K.; Ma, X.-C.; Chen, X.; Xue, Q.-K. Molecular Beam Epitaxy Growth of Superconducting LiFeAs Film on SrTiO$_3$ (001) Substrate. *EPL* **2015**, *109* (2), 28003. https://doi.org/10.1209/0295-5075/109/28003.

(22) Xu, J.-P.; Wang, M.-X.; Liu, Z. L.; Ge, J.-F.; Yang, X.; Liu, C.; Xu, Z. A.; Guan, D.; Gao, C. L.; Qian, D.; Liu, Y.; Wang, Q.-H.; Zhang, F.-C.; Xue, Q.-K.; Jia, J.-F. Experimental Detection of a Majorana Mode in the Core of a Magnetic Vortex inside a Topological Insulator-Superconductor Bi$_2$Te$_3$/NbSe$_2$ Heterostructure. *Phys. Rev. Lett.* **2015**, *114* (1), 017001. https://doi.org/10.1103/PhysRevLett.114.017001.

(23) Sun, H.-H.; Zhang, K.-W.; Hu, L.-H.; Li, C.; Wang, G.-Y.; Ma, H.-Y.; Xu, Z.-A.; Gao, C.-L.; Guan, D.-D.; Li, Y.-Y.; Liu, C.; Qian, D.; Zhou, Y.; Fu, L.; Li, S.-C.; Zhang, F.-C.; Jia, J.-F. Majorana Zero Mode Detected with Spin Selective Andreev Reflection in the Vortex of a Topological Superconductor. *Phys. Rev. Lett.* **2016**, *116* (25), 257003. https://doi.org/10.1103/PhysRevLett.116.257003.

(24) Trang, C. X.; Shimamura, N.; Nakayama, K.; Souma, S.; Sugawara, K.; Watanabe, I.; Yamauchi, K.; Oguchi, T.; Segawa, K.; Takahashi, T.; Ando, Y.; Sato, T. Conversion of a Conventional Superconductor into a Topological Superconductor by Topological Proximity Effect. *Nat Commun* **2020**, *11* (1), 159. https://doi.org/10.1038/s41467-019-13946-0.

(25) Haviland, D. B.; Liu, Y.; Goldman, A. M. Onset of Superconductivity in the Two-Dimensional Limit. *Phys. Rev. Lett.* **1989**, *62* (18), 2180–2183. https://doi.org/10.1103/PhysRevLett.62.2180.

(26) Yao, X.; Brahlek, M.; Yi, H. T.; Jain, D.; Mazza, A. R.; Han, M.-G.; Oh, S. Hybrid Symmetry Epitaxy of the Superconducting Fe(Te,Se) Film on a Topological Insulator. *Nano Lett.* **2021**, *21* (15), 6518–6524. https://doi.org/10.1021/acs.nanolett.1c01703.

(27) Trang, C. X.; Shimamura, N.; Nakayama, K.; Souma, S.; Sugawara, K.; Watanabe, I.; Yamauchi, K.; Oguchi, T.; Segawa, K.; Takahashi, T.; Ando, Y.; Sato, T. Conversion of a Conventional Superconductor into a Topological Superconductor by Topological Proximity Effect. *Nat Commun* **2020**, *11* (1), 159. https://doi.org/10.1038/s41467-019-13946-0.

(28) Qin, H.; Guo, B.; Wang, L.; Zhang, M.; Xu, B.; Shi, K.; Pan, T.; Zhou, L.; Chen, J.; Qiu, Y.; Xi, B.; Sou, I. K.; Yu, D.; Chen, W.-Q.; He, H.; Ye, F.; Mei, J.-W.; Wang, G. Superconductivity in Single-Quintuple-Layer $Bi_2Te_3$ Grown on Epitaxial FeTe. *Nano Lett.* **2020**, *20* (5), 3160–3168. https://doi.org/10.1021/acs.nanolett.9b05167.

(29) Lin, L.; Tian, Y.; Yu, W.; Chen, S.; Chen, Y.; Chen, W. Corrosion and Hardness Characteristics of Ti/TiN-Modified $Ti_6Al_4V$ Alloy in Marine Environment. *Ceram. Int.* **2022**, *48* (23), 34848–34854. https://doi.org/10.1016/j.ceramint.2022.08.074.

(30) Rodrigues, C. A. D.; Bandeira, R. M.; Duarte, B. B.; Tremiliosi-Filho, G.; Roche, V.; Jorge, A. M. Effect of Titanium Nitride (TiN) on the Corrosion Behavior of a Supermartensitic Stainless Steel. *Mater. & Corros.* **2019**, *70* (1), 28–36. https://doi.org/10.1002/maco.201810289.

(31) Silva, F. C.; Prada Ramirez, O. M.; Sagás, J. C.; Fontana, L. C.; De Melo, H. G.; Schön, C. G.; Tunes, M. A. High-Performance Titanium Nitride Structural Coatings for Corrosion Protection of Aluminum-Based Proton Exchange Membrane Fuel Cells. *ACS Materials Lett.* **2024**, *6* (10), 4564–4570. https://doi.org/10.1021/acsmaterialslett.4c01303.

(32) Ho, I. H.; Chang, C.-W.; Chen, Y.-L.; Chang, W.-Y.; Kuo, T.-J.; Lu, Y.-J.; Gwo, S.; Ahn, H. Ultrathin TiN Epitaxial Films as Transparent Conductive Electrodes. *ACS Appl. Mater. Interfaces* **2022**, *14* (14), 16839–16845. https://doi.org/10.1021/acsami.2c00508.

(33) Huang, Y.; Zhang, L.; Zhou, X.; Liao, L.; Jin, F.; Han, X.; Dong, T.; Xu, S.; Zhao, L.; Dai, Y.; Cheng, Q.; Huang, X.; Zhang, Q.; Wang, L.; Wang, N.-L.; Yue, M.; Bai, X.; Li, Y.; Wu, Q.; Gao, H.-J.; Gu, G.; Wang, Y.; Zhou, X.-J. Unveiling the Degradation Mechanism of High-Temperature

Superconductor $Bi_2Sr_2CaCu_2O_{8+\delta}$ in Water-Bearing Environments. *ACS Appl. Mater. Interfaces* **2022**, *14* (34), 39489–39496. https://doi.org/10.1021/acsami.2c08997.

(34) Ding, X.; Shen, S.; Leng, H.; Xu, M.; Zhao, Y.; Zhao, J.; Sui, X.; Wu, X.; Xiao, H.; Zu, X.; Huang, B.; Luo, H.; Yu, P.; Qiao, L. Stability of Superconducting $Nd_{0.8}Sr_{0.2}NiO_2$ Thin Films. *Sci. China Phys. Mech. Astron.* **2022**, *65* (6), 267411. https://doi.org/10.1007/s11433-021-1871-x.

(35) Cao, Y.; Mishchenko, A.; Yu, G. L.; Khestanova, E.; Rooney, A. P.; Prestat, E.; Kretinin, A. V.; Blake, P.; Shalom, M. B.; Woods, C.; Chapman, J.; Balakrishnan, G.; Grigorieva, I. V.; Novoselov, K. S.; Piot, B. A.; Potemski, M.; Watanabe, K.; Taniguchi, T.; Haigh, S. J.; Geim, A. K.; Gorbachev, R. V. Quality Heterostructures from Two-Dimensional Crystals Unstable in Air by Their Assembly in Inert Atmosphere. *Nano Lett.* **2015**, *15* (8), 4914–4921. https://doi.org/10.1021/acs.nanolett.5b00648.

(36) Pomjakushina, E.; Conder, K.; Pomjakushin, V.; Bendele, M.; Khasanov, R. Synthesis, Crystal Structure, and Chemical Stability of the Superconductor $FeSe_{1-x}$. *Phys. Rev. B* **2009**, *80* (2), 024517. https://doi.org/10.1103/PhysRevB.80.024517.

(37) Xie, R.; Ge, M.; Xiao, S.; Zhang, J.; Bi, J.; Yuan, X.; Yi, H. T.; Wang, B.; Oh, S.; Cao, Y.; Yao, X. Resilient Growth of a Highly Crystalline Topological Insulator–Superconductor Heterostructure Enabled by an *Ex Situ* Nitride Film. *ACS Appl. Mater. Interfaces* **2024**, *16* (26), 34386–34392. https://doi.org/10.1021/acsami.4c05656.

(38) Guo, Z.; Ge, M.; Zhou, Y.-Q.; Bi, J.; Zhang, Q.; Zhang, J.; Ye, J.-T.; Zhai, R.; Ge, F.; Huang, Y.; Zhang, R.; Yao, X.; Huang, L.-F.; Cao, Y. High Resistance of Superconducting TiN Thin Films against Environmental Attacks. *Mater. Horiz.* **2024**, *11* (23), 5972–5982. https://doi.org/10.1039/D4MH00959B.

(39) Bi, J.; Lin, Y.; Zhang, Q.; Liu, Z.; Zhang, Z.; Zhang, R.; Yao, X.; Chen, G.; Liu, H.; Huang, Y.; Sun, Y.; Zhang, H.; Sun, Z.; Xiao, S.; Cao, Y. Momentum-Resolved Electronic Structures and Strong Electronic Correlations in Graphene-like Nitride Superconductors. *Nano Lett.* **2024**, *24* (24), 7451–7457. https://doi.org/10.1021/acs.nanolett.4c01704.

(40) Zhou, Y.-Q.; Guo, Z.; Bi, J.; Ye, J.-T.; Ge, M.; Lin, Y.; Xiao, S.; Cao, Y.; Wang, L.; Huang, L.-F. Unifying the Atomistic Trends for Early-Stage Evolution of TiN Surfaces in Atmospheric and Aqueous Environments. *Acta Mater* **2025**, *289*, 120909. https://doi.org/10.1016/j.actamat.2025.120909.

(41) Clarke, J. The Proximity Effect between Superconducting and Normal Thin Films in Zero Field. *J. Phys. Colloques* **1968**, *29* (C2), C2-3-C2-16. https://doi.org/10.1051/jphyscol:1968201.

(42) Hilsch, P. Zum Verhalten von Supraleitern im Kontakt mit Normalleitern. *Z. Physik* **1962**, *167*, 511–524. https://doi.org/10.1007/bf01378178.
22

(43) Zhang, H.; Liu, C.-X.; Qi, X.-L.; Dai, X.; Fang, Z.; Zhang, S.-C. Topological Insulators in $Bi_2Se_3$, $Bi_2Te_3$ and $Sb_2Te_3$ with a Single Dirac Cone on the Surface. *Nat Phys* **2009**, *5* (6), 438–442. https://doi.org/10.1038/nphys1270.

(44) Chen, Y. L.; Analytis, J. G.; Chu, J.-H.; Liu, Z. K.; Mo, S.-K.; Qi, X. L.; Zhang, H. J.; Lu, D. H.; Dai, X.; Fang, Z.; Zhang, S. C.; Fisher, I. R.; Hussain, Z.; Shen, Z.-X. Experimental Realization of a Three-Dimensional Topological Insulator, $Bi_2Te_3$. *Science* **2009**, *325* (5937), 178–181. https://doi.org/10.1126/science.1173034.

(45) Li, Y.; Wang, G.; Zhu, X.; Liu, M.; Ye, C.; Chen, X.; Wang, Y.; He, K.; Wang, L.; Ma, X.; Zhang, H.; Dai, X.; Fang, Z.; Xie, X.; Liu, Y.; Qi, X.; Jia, J.; Zhang, S.; Xue, Q. Intrinsic Topological Insulator $Bi_2Te_3$ Thin Films on Si and Their Thickness Limit. *Adv. Mater.* **2010**, *22* (36), 4002–4007. https://doi.org/10.1002/adma.201000368.

(46) Wang, B. Y.; Li, D.; Goodge, B. H.; Lee, K.; Osada, M.; Harvey, S. P.; Kourkoutis, L. F.; Beasley, M. R.; Hwang, H. Y. Isotropic Pauli-Limited Superconductivity in the Infinite-Layer Nickelate $Nd_{0.775}Sr_{0.225}NiO_2$. *Nat. Phys.* **2021**, *17* (4), 473–477. https://doi.org/10.1038/s41567-020-01128-5.

(47) Ma, K.; Gornicka, K.; Lefèvre, R.; Yang, Y.; Rønnow, H. M.; Jeschke, H. O.; Klimczuk, T.; Von Rohr, F. O. Superconductivity with High Upper Critical Field in the Cubic Centrosymmetric η-Carbide $Nb_4Rh_2C_{1-\delta}$. *ACS Mater. Au* **2021**, *1* (1), 55–61. https://doi.org/10.1021/acsmaterialsau.1c00011.

(48) Ruggiero, S. T.; Barbee, T. W.; Beasley, M. R. Superconductivity in Quasi-Two-Dimensional Layered Composites. *Phys. Rev. Lett.* **1980**, *45* (15), 1299–1302. https://doi.org/10.1103/PhysRevLett.45.1299.

(49) Kresse, G.; Hafner, J. *Ab Initio* Molecular Dynamics for Liquid Metals. *Phys. Rev. B* **1993**, *47* (1), 558–561. https://doi.org/10.1103/PhysRevB.47.558.

(50) Kresse, G.; Furthmüller, J. Efficient Iterative Schemes for *Ab Initio* Total-Energy Calculations Using a Plane-Wave Basis Set. *Phys. Rev. B* **1996**, *54* (16), 11169–11186. https://doi.org/10.1103/PhysRevB.54.11169.

(51) Perdew, J. P.; Burke, K.; Ernzerhof, M. Generalized Gradient Approximation Made Simple. *Phys. Rev. Lett.* **1996**, *77* (18), 3865–3868. https://doi.org/10.1103/PhysRevLett.77.3865.

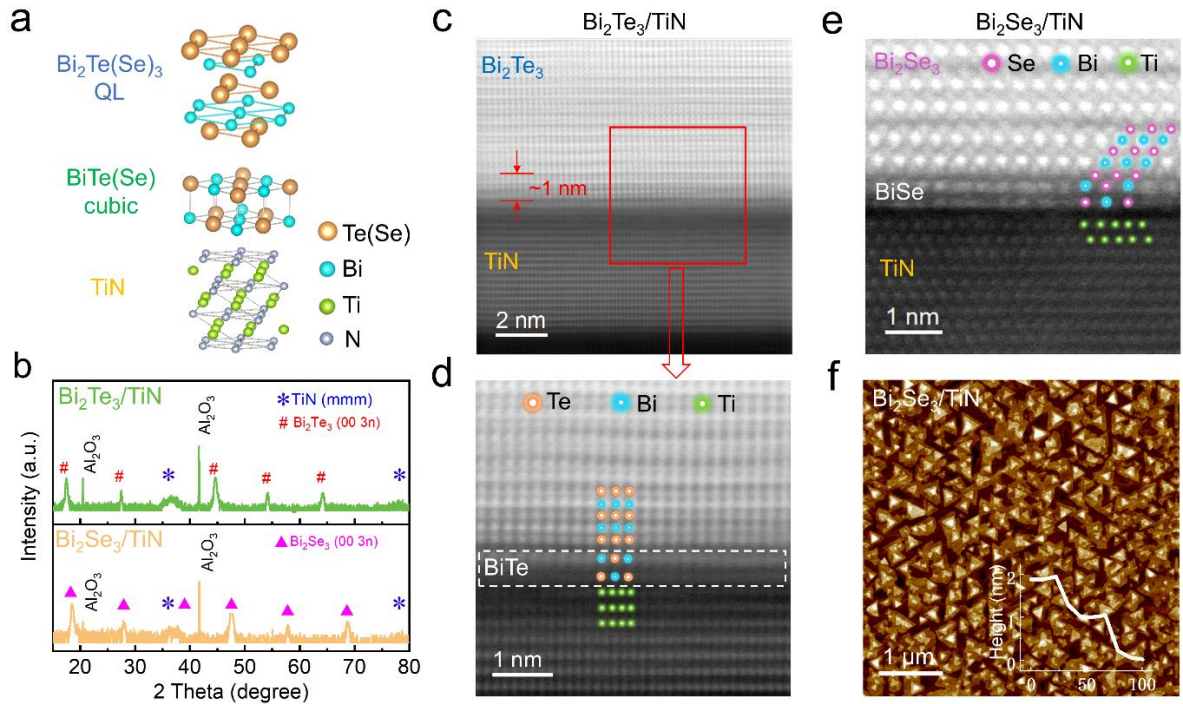

**Figure 1. Structural characterizations of Bi$_2$Te$_3$(Bi$_2$Se$_3$)/TiN heterostructures.** (a) Schematic illustration of the Bi$_2$Te$_3$(Bi$_2$Se$_3$)/TiN heterostructure. (b) XRD profiles of 25nm Bi$_2$Te$_3$/TiN/Al$_2$O$_3$ and 50nm Bi$_2$Se$_3$/TiN/Al$_2$O$_3$. The marks labeled Bi$_2$Te$_3$ (00 3n), Bi$_2$Se$_3$ (00 3n), and TiN peaks. (c) STEM-HAADF image of the Bi$_2$Se$_3$/TiN heterostructure. The red arrows indicate the thickness of 1 QL is about 1 nm. (d) An enlarged image of (c), a cubic BiTe bilayer, is clearly visible at the interface. (e) STEM-HAADF image of the Bi$_2$Se$_3$/TiN heterostructure. A similar BiSe bilayer is also observed at the interface. (f) AFM morphology of a 25nm Bi$_2$Se$_3$/TiN/Al$_2$O$_3$ sample. The inset shows the height profile of the terraces with step height ~ 1nm, corresponding to 1 QL of Bi$_2$Se$_3$.

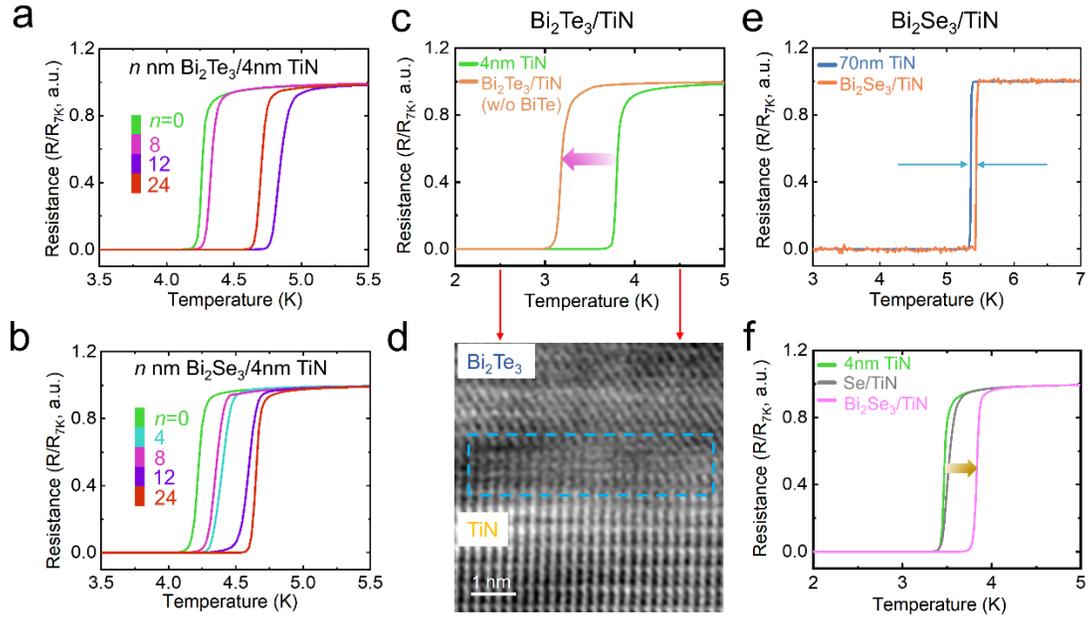

**Figure 2. Superconductivity enhancement in $Bi_2Te_3$($Bi_2Se_3$)/TiN heterostructures.** R-T curves of (a) n nm $Bi_2Te_3$/4nm TiN and (b) n nm $Bi_2Se_3$/4nm TiN heterostructures from 3.5 K to 5.5 K under zero magnetic field. (c) R-T curves of 4nm TiN and 12nm $Bi_2Te_3$/4nm TiN with a deteriorated interface. The $Bi_2Te_3$/TiN heterostructure shows a $T_c$ reduction of 0.6 K compared with pure TiN. (d) STEM-HAADF image of the $Bi_2Te_3$/TiN heterostructure in (c). The blue dashed box shows a deteriorated interface without BiTe bilayer intercalation. (e) R-T curves of 70 nm TiN and 50 nm $Bi_2Se_3$/70nm TiN. The $T_c$ of the two samples is nearly identical (change <0.1 K). (f) R-T curves of 4nm TiN, 140nm Se/4nm TiN, and 50nm $Bi_2Se_3$/4nm TiN. The $T_c$ of Se/TiN is the same of pure TiN film, while the $Bi_2Se_3$/TiN heterostructure shows an enhancement of 0.4 K. In all the above images, R is normalized by the normal state resistance at T = 7.0 K. All samples were grown on $Al_2O_3$ substrates except for the films in Figure 2b, which were grown on $KTaO_3$ substrates.

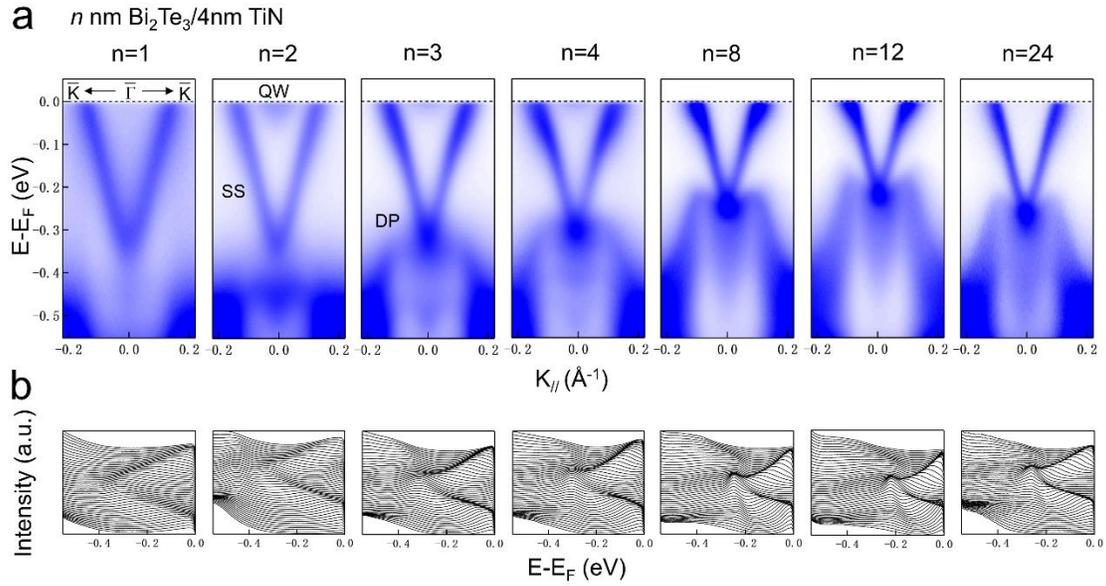

**Figure 3. ARPES band spectra of n nm Bi₂Te₃/4nm TiN heterostructures.** (a) ARPES results with n=1, 2, 3, 4, 8, 12, 24. (b) Corresponding energy distribution curves of ARPES results in (a). All spectra were collected along the $\overline{\Gamma}-\overline{K}$ direction. The bulk quantum well appears at n=2 and disappears at n=8. The surface state appears at n≥2, and the Dirac point is formed at n=3. All the data were collected at 7.3K.

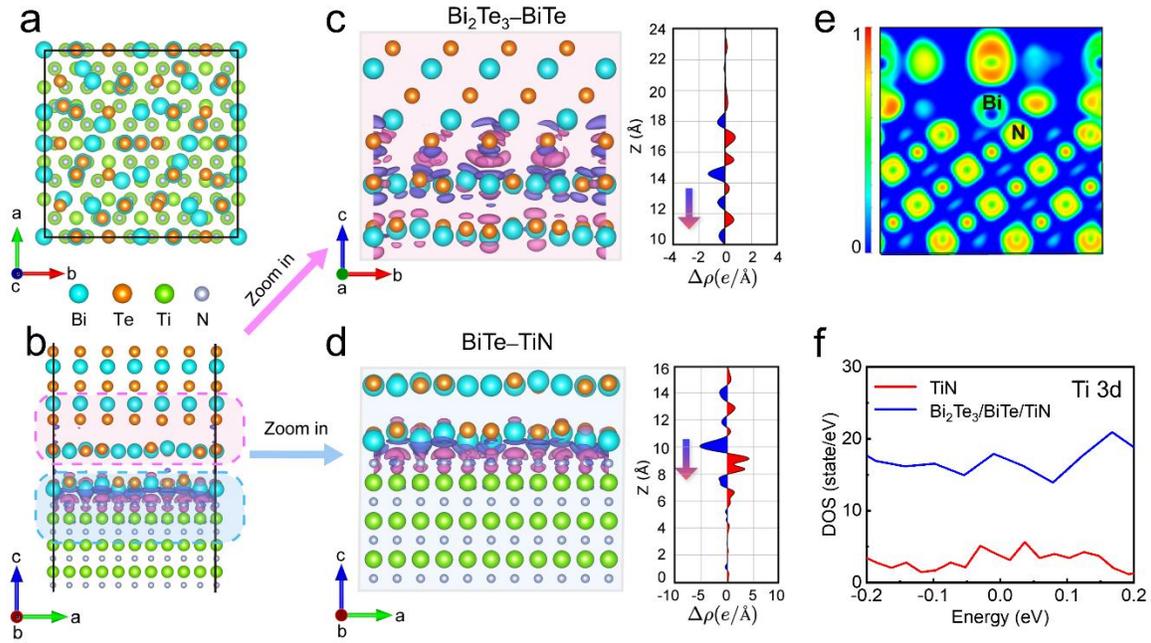

**Figure 4. Theoretical calculations on Bi$_2$Te$_3$/BiTe/TiN heterostructure.** (a-b) Top and side views of the Bi$_2$Te$_3$/BiTe/TiN heterostructure, respectively. (c-d) Differential charge density distributions at the Bi$_2$Te$_3$–BiTe (isosurface level = 0.002 e/Å) and BiTe–TiN (isosurface level = 0.009 e/Å) interfaces (left), along with the corresponding integrals of the differential charge densities (right). Positive isosurface is indicated in red, and the negative counterpart in blue, at a fixed isosurface level. (e) Electron localization function (ELF) at the BiTe–TiN interface. (f) Ti atom-resolved partial density of states (DOS) of TiN and the Bi$_2$Te$_3$/BiTe/TiN heterostructure.

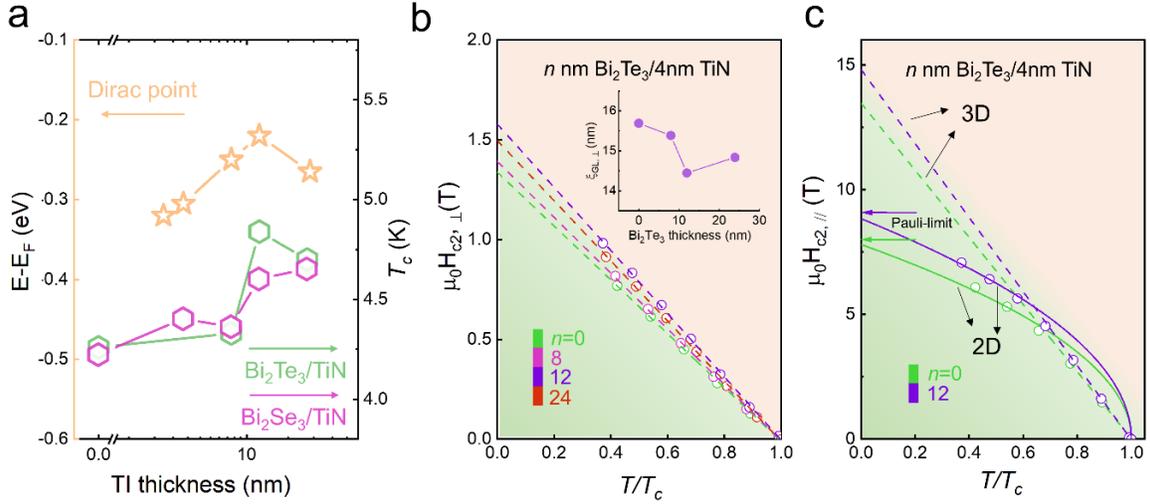

**Figure 5. $T_c$ and upper critical fields of Bi$_2$Te$_3$(Bi$_2$Se$_3$)/TiN heterostructures.** (a) Thickness dependence of the Dirac point (Bi$_2$Te$_3$ ⩾ 3 nm) and the $T_c$ of Bi$_2$Te$_3$/TiN (Bi$_2$Se$_3$/TiN) heterostructures. The upper critical fields (H$_{c2,\perp}$ and H$_{c2,//}$) of the Bi$_2$Te$_3$/TiN heterostructures with magnetic fields applied perpendicular (b) and parallel (c) to the ab plane. The inset in (b) shows the out-of-plane GL coherence lengths as a function of Bi$_2$Te$_3$ thickness. The dash and solid lines in (b) represent fitting results by 3D and 2D GL theory, respectively. The green and purple arrows indicate the Pauli-limit of H$_{c2,//}$ for TiN and Bi$_2$Te$_3$/TiN, which are 8.0 T and 9.1 T, respectively.

**For TOC only**

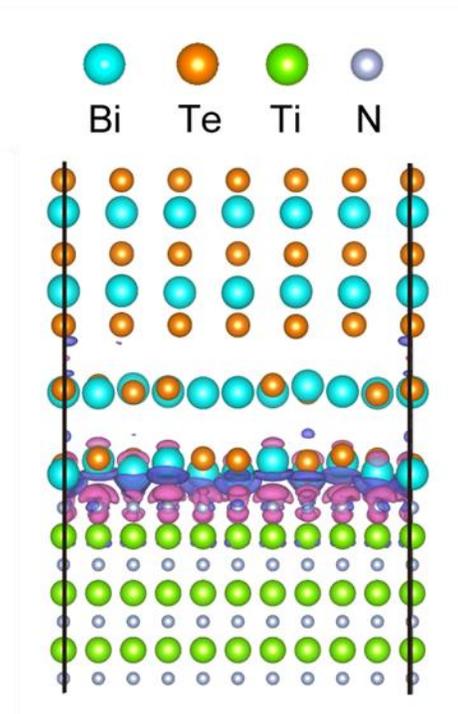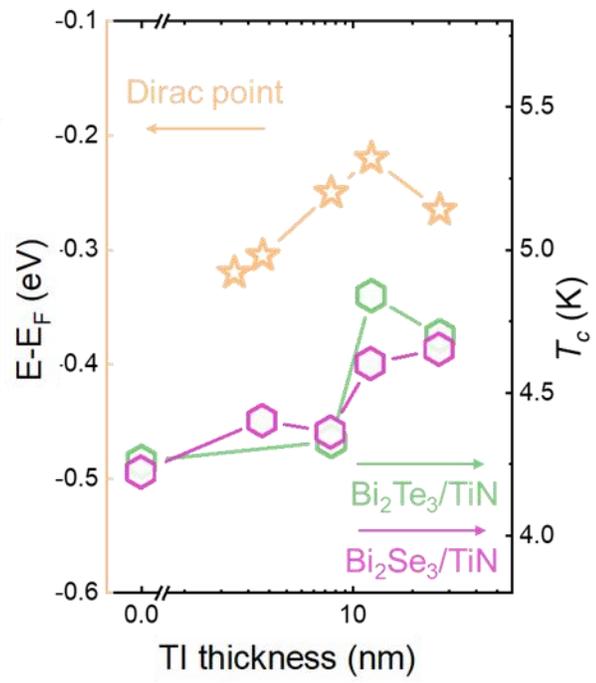